# A New Method for Constructing Large Size WBE Codes with Low Complexity ML Decoder

Mohammad Javad Faraji, Pedram Pad, and Farokh Marvasti

Advanced Communications Research Institute (ACRI)
Department of Electrical Engineering, Sharif University of Technology, Tehran, Iran
Email: {faraji, pedram_pad}@ee.sharif.edu, marvasti@sharif.edu

*Abstract*—In this paper we wish to introduce a method to reconstruct large size Welch Bound Equality (WBE) codes from small size WBE codes. The advantage of these codes is that the implementation of ML decoder for the large size codes is reduced to implementation of ML decoder for the core codes. This leads to a drastic reduction of the computational cost of ML decoder. Our method can also be used for constructing large Binary WBE (BWBE) codes from smaller ones. Additionally, we explain that although WBE codes are maximizing the sum channel capacity when the inputs are real valued, they are not necessarily appropriate when the input alphabet is binary. The discussion shows that when the input alphabet is binary, the Total Squared Correlation (TSC) of codes is not a proper figure of merit.

## I. INTRODUCTION

In Direct Sequence Code Division Multiple Access (DS-CDMA) each user is assigned a signature vector for transmitting the data through a common channel (in time and frequency). As proved in [1] the sum channel capacity of such a system is maximized when the Total Squared Correlation (TSC) of the signature vectors meets its lower bound; i.e. the Welch bound [2]. These signature sets are called Welch Bound Equality (WBE) codes [3]. It is worth mentioning that the channel capacity is maximized when the input distribution of the user data are Gaussian. Some methods for constructing WBE codes are proposed in [4-6].

For practical conditions, it is favorable to use binary antipodal signatures. Limiting the signatures to binary antipodal vectors, signature sets that meet the Welch bound may not exist. For these codes, the lower bounds of TSC for different spreading factors ($L$) and different number of users ($K$) are derived in [7-8]. These bounds are known as Karystinos-Pados (KP) bounds and are presented in Tabs. 1 and 2. Notice that in the under-loaded case when $L$ is a multiple of 4, the KP bound is equal to the Welch bound. Furthermore, in the over-loaded case the KP bound is equal to the Welch bound when $K$ is a multiple of 4. Some methods for constructing binary antipodal signatures that meet KP bounds are proposed in [3], [7-10].

Clearly, to have high performance decoders for the over-loaded systems we should perform Multi User Detection (MUD) [11-13]. Some examples of the suggested decoding methods are PIC [14-15], SIC [16], and Iterative interference cancelation [17-18]. These methods are somewhat heuristic and none of them is proved to be optimum. Practical realization of optimum decoders has not been proposed yet.

In this paper, first we introduce a method for constructing large WBE codes by enlarging the smaller ones. We also prove that our method can be used for enlarging the binary antipodal codes when the KP bound meets the Welch bound. These codes have the advantage of low computational complexity optimum decoder. In fact, their decoding problem can be reduced to decoding problems of the core codes. This means that the Maximum Likelihood (ML) decoding of the smaller codes provide ML decoding of the large codes. This leads to a dramatic decrease in computational complexity. For example, using the WBE codes that are constructed by the proposed method, we simulate a CDMA system with 64 chips and 96 users with binary antipodal input alphabet and ML decoding. It is noticeable that the ML decoder of such system was not practically implementable until now (notice that binary input alphabet is encouraging from the practical point of view).

In the next section we propose a method for constructing large WBE codes from smaller ones. Section III includes the method of decoding the proposed codes. Special discussions about CDMA systems with binary antipodal input alphabets are done in section IV. Simulation results are discussed in section V. Section VI consists of conclusion and future works.

| Processing Gain | Number of Users | Lower Bound on TSC |
|---|---|---|
| $L \equiv 0 \pmod 4$ | Any $K$ | $K$ |
| $L \equiv 2 \pmod 4$ | $K \equiv 0 \pmod 2$ | $K + 2\frac{K(K-2)}{L^2}$ |
| $L \equiv 2 \pmod 4$ | $K \equiv 1 \pmod 2$ | $K + 2\left(\frac{K-1}{L}\right)^2$ |
| $L \equiv 1 \pmod 2$ | Any $K$ | $K + \frac{K(K-1)}{L^2}$ |

Tab. 1. Under-loaded DS-CDMA System ($K \leq L$) [7]

| Number of Users | Processing Gain | Lower Bound on TSC |
|---|---|---|
| $K \equiv 0 \pmod 4$ | Any $L$ | $\frac{K^2}{L}$ |
| $K \equiv 2 \pmod 4$ | $L \equiv 0 \pmod 2$ | $\frac{K^2}{L} + 2\frac{L-2}{L}$ |
| $K \equiv 2 \pmod 4$ | $L \equiv 1 \pmod 2$ | $\frac{K^2}{L} + 2\left(\frac{L-1}{L}\right)^2$ |
| $K \equiv 1 \pmod 2$ | Any $L$ | $\frac{K^2}{L} + \frac{L-1}{L}$ |

Tab. 2. Over-loaded DS-CDMA System ($K \geq L$) [7]

## II. NEW METHOD FOR CONSTRUCTING WBE CODES

A CDMA system in an AWGN channel can be modeled as
$$Y = CX + N \quad (1)$$
where $\mathbf{C}$ is the $L \times K$ code matrix (each of its columns is the signature of each user), $X$ is the $K \times 1$ vector of the user data, $N$ is a Gaussian noise vector with zero mean and auto-covariance matrix of $\sigma^2 \mathbf{I}_L$ (where $\mathbf{I}_L$ is the $L \times L$ identity matrix).

For maximizing the sum channel capacity of such a system, we should find codes which are as orthogonal as possible according to the TSC criterion [1]. As defined in [1]
$$\text{TSC}(\mathbf{C}) = \sum_{i=1}^{K} \sum_{j=1}^{K} |C_i^H \cdot C_j|^2 \quad (2)$$
where $C_i$ is the $i^{\text{th}}$ column of $\mathbf{C}$ which is normalized. As proved in [2]
$$\text{TSC}(\mathbf{C}) \geq \begin{cases} K & K \leq L \\ K^2/L & K \geq L \end{cases} \quad (3)$$
The codes that their TSC meet this lower bound are called WBE codes [3].

Using the obvious expression that
$$\text{TSC}(\mathbf{C}) = \sum_{i=1}^{K} \sum_{j=1}^{K} |d_{ij}|^2 \quad (4)$$
where $\mathbf{D} = [d_{ij}] = \mathbf{C}^H \mathbf{C}$, we construct large WBE codes from smaller ones in the following theorem.

**Theorem 1** If $\mathbf{C}$ is a $L \times K$ WBE matrix and $\mathbf{Q}$ is a $d \times d$ unitary matrix, then $\mathbf{Q} \otimes \mathbf{C}$ is a $dL \times dK$ WBE matrix, where $\otimes$ denotes the Kronecker product.

*Proof*: Let $\mathbf{D} = [d_{ij}] = \mathbf{C}^H \mathbf{C}$. We have
$$\mathbf{E} = [e_{ij}] = (\mathbf{Q} \otimes \mathbf{C})^H \cdot (\mathbf{Q} \otimes \mathbf{C}) = \mathbf{I}_d \otimes \mathbf{D} \quad (5)$$
According to (4)
$$\text{TSC}(\mathbf{Q} \otimes \mathbf{C}) = \sum_{i=1}^{dK} \sum_{j=1}^{dK} |e_{ij}|^2 \quad (6)$$
Using (5) and (6), we have
$$\text{TSC}(\mathbf{Q} \otimes \mathbf{C}) = d \cdot \sum_{i=1}^{K} \sum_{j=1}^{K} |d_{ij}|^2 = d \cdot \text{TSC}(\mathbf{C}) \quad (7)$$
According to (7), if $K \leq L$, then $\text{TSC}(\mathbf{C}) = K$ and $\text{TSC}(\mathbf{Q} \otimes \mathbf{C}) = dK$ which is equal to the Welch bound for the $dL \times dK$ matrix. If $K \geq L$, then $\text{TSC}(\mathbf{C}) = K^2/L$ and $\text{TSC}(\mathbf{Q} \otimes \mathbf{C}) = d K^2/L = (dK)^2/(dL)$ which is equal to the Welch bound for the $dL \times dK$ matrix. Therefore, $\mathbf{Q} \otimes \mathbf{C}$ is a WBE matrix. ∎

This theorem provides a systematic way for constructing large WBE codes from smaller ones.

**Example 1** In Theorem 1, if $\mathbf{Q}$ is the identity matrix, then the code matrix $\mathbf{Q} \otimes \mathbf{C}$ is equivalent to a multiple access channel with $d$ TDMA channels, each of these channels consisting of $K$ CDMA channels.

**Corollary 1** In Theorem 1, if $\mathbf{C}$ is a binary antipodal matrix and $L$ and $K$ are in the mode that the KP bound equals the Welch bound, then $\left(\frac{1}{\sqrt{d}} \mathbf{H}_d\right) \otimes \mathbf{C}$ is a $dL \times dK$ binary antipodal matrix where $\mathbf{H}_d$ is a $d \times d$ Hadamard matrix.

In the next section we will propose a method for reducing the decoding problem of the large codes constructed in Theorem 1 to the decoding problem of the smaller codes.

## III. THE DECODING OF THE PROPOSED CODES

In this section we use the special structure of the proposed large codes to reduce their decoding problem to the decoding problem of the smaller codes.

Suppose $\mathbf{C}$ is a $L \times K$ matrix and $\mathbf{Q}$ is a $d \times d$ unitary matrix. Similar to (1) in an AWGN channel the received vector is
$$Y = (\mathbf{Q} \otimes \mathbf{C})X + N \quad (8)$$
In which $X$, $N$ and $Y$ are $dK \times 1$, $dL \times 1$ and $dL \times 1$ vectors, respectively. Multiplying both sides by $\mathbf{Q}^H \otimes \mathbf{I}_L$, we have
$$Y' = (\mathbf{Q}^H \otimes \mathbf{I}_L)Y = (\mathbf{Q}^H \otimes \mathbf{I}_L)(\mathbf{Q} \otimes \mathbf{C})X + N'$$
$$= (\mathbf{I}_d \otimes \mathbf{C})X + N' \quad (9)$$
where the subscript of $\mathbf{I}$ determines the dimension of the identity matrix and $N' = (\mathbf{Q}^H \otimes \mathbf{I}_L)N$. Obviously,
$$(\mathbf{Q}^H \otimes \mathbf{I}_L)^H \cdot (\mathbf{Q}^H \otimes \mathbf{I}_L) = (\mathbf{Q} \otimes \mathbf{I}_L) \cdot (\mathbf{Q}^H \otimes \mathbf{I}_L) = \mathbf{I}_{dL} \quad (10)$$
and thus $\mathbf{Q}^H \otimes \mathbf{I}_L$ is a unitary matrix. Since $\mathbf{Q}^H \otimes \mathbf{I}_L$ is a unitary matrix and $N$ is a Gaussian random vector with zero mean and auto-covariance matrix $\sigma^2 \mathbf{I}_{dL}$, $N'$ is a random vector with properties identical to $N$. In other words, the entries of $N'$ are independent Gaussian random variable with zero mean and variance $\sigma^2$. Hence, the ML extraction of the $X$ vector from $Y$ is equivalent to ML extraction of the $X$ vector from $Y'$.

Rewriting (9), we have
$$Y' = \begin{bmatrix} \mathbf{C} & \cdots & 0 \\ \vdots & \ddots & \vdots \\ 0 & \cdots & \mathbf{C} \end{bmatrix} X + N' \quad (11)$$
This means that the first $L$ entries of $Y'$ depend only on the first $K$ entries of $X$ and first $L$ entries of $N'$, the second $L$ entries of $Y'$ depend only on the second $K$ entries of $X$ and second $L$ entries of $N'$, and so on. In other words, the problem of detecting $X$ from the received vector is decoupled to $d$ smaller problems. This leads to a dramatic reduction in computational complexity in overloaded systems ($K \geq L$).

In the remaining of this paper, we focus on the CDMA systems with binary antipodal input alphabet i.e., from now on we assume that the data vector $X$ in (1) belongs to the set $\{-1, +1\}^K$.

**Corollary 2** Suppose that we have a CDMA system with $dL$ chips, $dK$ users and WBE signatures. For direct implementation of the ML decoder, we need to calculate about $2^{dK}$ Euclidean distances. But if we used the WBE signatures that have been proposed in the previous section, for the decoding problem we need to calculate only $d \times 2^K$ Euclidean distance. This means a tremendous reduction in the complexity of the decoder. Some similar systems are simulated in section V.

## IV. SPECIAL DISCUSSIONS ABOUT CDMA SYSTEMS WITH BINARY ANTIPODAL INPUT ALPHABET

We know that the direct implementation of ML decoder from Bit Error Rate (BER) point of view needs a large amount of computational complexity. In the following, we desire to propose a decoding method with very low computational complexity which is, under some conditions, ML from Symbol Error Rate (SER) point of view. We call this method SER Almost ML (AML) decoding. Although in the over-loaded case the SER ML decoding is not equivalent to BER

ML decoding, the SER optimum decoder is somewhat suboptimum from BER point of view. It will be seen in the next section, despite the fact that SER AML decoding has significantly lower computational complexity than BER ML, they have very similar performance in slightly high $E_b/N_0$.

Performing the ML decoder for an AWGN channel from SER point of view, we have

$$\hat{X} = \underset{X \in \{-1,+1\}^K}{\operatorname{argmin}} \|Y - \mathbf{C}X\| \quad (12)$$

where $\| \|$ indicates the Euclidean norm.

Direct implementation of this method needs about $2^K$ searches which is not practical for usual $K$'s. Now suppose that $\mathbf{C}$ is full rank. By permuting the columns of $\mathbf{C}$ we can write $\mathbf{C} = [\mathbf{A}|\mathbf{B}]$ where $\mathbf{A}$ in an $L \times L$ invertible matrix. Rewriting (1), we have

$$Y = \mathbf{C}X + N = \mathbf{A}X_1 + \mathbf{B}X_2 + N \quad (13)$$

where $X_1$ and $X_2$ are $L \times 1$ and $(K - L) \times 1$ vectors, respectively. Now multiplying both sides by $\mathbf{A}^{-1}$, we arrive at

$$\mathbf{A}^{-1}Y = X_1 + \mathbf{A}^{-1}\mathbf{B}X_2 + N' \quad (14)$$

where $N' = \mathbf{A}^{-1}N$. Similar to (12), we are looking for

$$\check{X} = \begin{bmatrix} \check{X}_1 \\ \check{X}_2 \end{bmatrix} = \underset{\substack{X_1 \in \{-1,+1\}^L \\ X_2 \in \{-1,+1\}^{K-L}}}{\operatorname{argmin}} \|\mathbf{A}^{-1}Y - X_1 - \mathbf{A}^{-1}\mathbf{B}X_2\|$$

(15)

But it is clear that the nearest $\{-1,+1\}$-vector to a vector $Z$ is $sign(Z)$ which is obtained by substituting the positive entries of $Z$ with $+1$ and its negative entries with $-1$. Using this fact, we find that

$$\check{X}_2 = \underset{X_2 \in \{-1,+1\}^{K-L}}{\operatorname{argmin}} \|(\mathbf{A}^{-1}Y - \mathbf{A}^{-1}\mathbf{B}X_2) - sign(\mathbf{A}^{-1}Y - \mathbf{A}^{-1}\mathbf{B}X_2)\| \quad (16)$$

and

$$\check{X}_1 = sign(\mathbf{A}^{-1}Y - \mathbf{A}^{-1}\mathbf{B}\check{X}_2) \quad (17)$$

Having a precise look at (16) and (17), we discover that it only needs to search among $2^{K-L}$ vectors rather than $2^K$ vectors. It means a significant decrease in the complexity of the decoder. Now notice that if $\mathbf{A}$ is a unitary matrix, then $N'$ is a noise vector with the same properties as $N$ and thus $\check{X} = \hat{X}$. This means that by following (16) and (17) we can implement the SER ML decoder with a much lower computational complexity than the direct implementation of the SER ML decoder. However, we call the decoder that is obtained by (16) and (17) Almost ML (AML) decoder.

It is worth mentioning that if the multiplication of the matrix $\mathbf{C}$ with $\{-1, +1\}$-vectors is not one-to-one, then there are always cases that $\hat{X}$ and $\check{X}$ are not unique. This leads to an interference that cannot be removed and an intrinsic non-zero probability of error. Considering this point, we will show through simulations that despite the fact that the codes with minimum TSC maximize the channel capacity when the input data are real or complex, they are not necessarily appropriate for a CDMA system with binary inputs. In other words, we will show that the TSC of a code is not a proper criterion for comparing the performance of the codes when the input alphabet is binary. Particularly, we will see WBE codes with very high noiseless BER and two codes with the same TSC and different performances. These phenomena bring this idea to mind that when the input alphabet is binary the main criterion that should be considered for designing a code is the extent to which it shows injective properties.

Simulation results for verifying the discussed facts are covered in the next section.

## V. SIMULATION RESULTS

To verify the previous discussions, we simulated four binary input CDMA systems with different binary antipodal signature matrices and different decoding schemes. In the following, $\mathbf{C}_{L \times K}$ and $\mathbf{C}'_{L \times K}$ denote different $L \times K$ BWBE codes which are constructed through the flowchart that is depicted in [7]. In addition, $\mathbf{K}_d = \frac{1}{\sqrt{d}}\mathbf{H}_d$ where $\mathbf{H}_d$ denotes a $d \times d$ Hadamard matrix.

**Simulation 1** The first system has a spreading factor of 56 with 64 users. According to Tab. 2 and equation (3), when $L = 7$ and $K = 8$ the KP bound equals the Welch bound. Thus, according to Theorem 1, $\mathbf{K}_8 \otimes \mathbf{C}_{7 \times 8}$ is a $56 \times 64$ BWBE code. The performance curves for this code, using BER ML, SER AML and Iterative interference cancellation with soft-limiter (IT) are depicted in Fig. 1. Additionally, the performance curve for $\mathbf{C}_{56 \times 64}$ with IT decoder is derived.

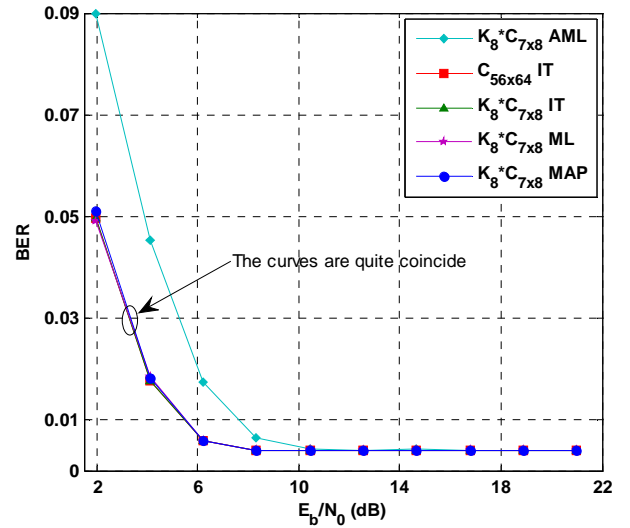

Fig. 1. BER versus $E_b/N_0$ for a system with 56 chips and 64 users using different decoding methods. In the figure, * indicates the Kronecker product.

As it is clear from Fig. 1 there is a significant amount of error that cannot be removed even by increasing $E_b/N_0$ (notice that the optimum decoder is performed).

**Simulation 2** The second system has a spreading factor of 64 with 72 users. According to Tab. 2 and equation (3), when $L = 8$ and $K = 9$ the KP bound is greater than the Welch bound. Thus, according to (7), $\mathbf{K}_8 \otimes \mathbf{C}_{8 \times 9}$ and $\mathbf{K}_8 \otimes \mathbf{C}'_{8 \times 9}$ are not BWBE codes but have the same TSCs which is very near to the KP bound. We call such codes Almost BWBE (ABWBE). The advantage of these codes is that their optimum and suboptimum decoder can be implemented just through the method proposed in section III and IV. The performance curves of $\mathbf{K}_8 \otimes \mathbf{C}_{8 \times 9}$, $\mathbf{K}_8 \otimes \mathbf{C}'_{8 \times 9}$ and $\mathbf{C}_{64 \times 72}$ using different decoding methods are depicted in Fig. 2.

Some interesting phenomena can be seen through this simulation. The first interesting point is that, despite the fact that the TSC of $\mathbf{K}_8 \otimes \mathbf{C}_{8\times 9}$ and $\mathbf{K}_8 \otimes \mathbf{C}'_{8\times 9}$ are equal, they have completely different performances. Using $\mathbf{K}_8 \otimes \mathbf{C}_{8\times 9}$ the BER tends to 0 when $E_b/N_0$ increases. However, when we use $\mathbf{K}_8 \otimes \mathbf{C}'_{8\times 9}$ the BER cannot be lower than a specific value. The second point is that, although the BWBE matrix $\mathbf{C}_{64\times 72}$ has better performance in low values of $E_b/N_0$, its BER curve saturates and is above the BER curve of the ABWBE $\mathbf{K}_8 \otimes \mathbf{C}_{8\times 9}$ matrix. Although some amount of the error of $\mathbf{C}_{64\times 72}$ may be introduced through the non-optimum IT decoder, it has an intrinsic amount of error because of the non-invertibility of its mapping on $\{-1, +1\}^{72}$. Notice that the SER AML decoder of $\mathbf{C}_{8\times 9}$ is equivalent to its SER ML decoder because its first 8 columns is a unitary matrix.

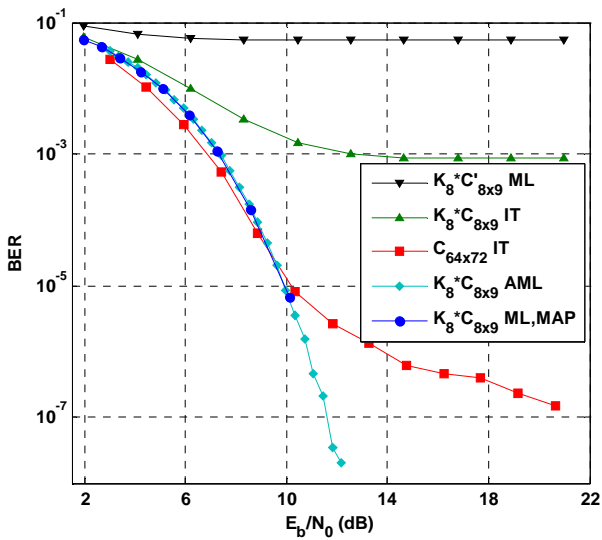

Fig. 2. BER versus $E_b/N_0$ for a system with 64 chips and 72 users using different codes and decoding methods. In the figure, * indicates the Kronecker product.

**Simulation 3** The third simulation is a system with 64 chips and 96 users. The simulations are done for $\mathbf{K}_8 \otimes \mathbf{C}_{8\times 12}$ and $\mathbf{C}_{64\times 96}$ with different decoding schemes. Notice that according to Tab. 2, equation (3) and Theorem 1, $\mathbf{K}_8 \otimes \mathbf{C}_{8\times 12}$ is a $64 \times 96$ BWBE code. The results are depicted in Fig. 3.

It is surprising to see that in this case random codes perform better than WBE codes in high $E_b/N_0$. This again shows that being WBE and having low TSC is not a proper criterion when the input alphabet is binary.

**Simulation 4** The last simulation is a system with 64 chips and 104 users. Again by referring to Tab. 2 and equation (3), we find that $\mathbf{K}_8 \otimes \mathbf{C}_{8\times 13}$ and $\mathbf{K}_8 \otimes \mathbf{C}'_{8\times 13}$ are not BWBE codes.

The contradicting point in this simulation is the different performance of $\mathbf{K}_8 \otimes \mathbf{C}_{8\times 13}$ and $\mathbf{K}_8 \otimes \mathbf{C}'_{8\times 13}$ that have the same TSCs. However, in this case the BWBE codes perform better than the others.

## VI. CONCLUSION AND FUTURE WORKS

In this paper we introduced a new method for constructing large WBE codes from smaller ones. The advantage of these codes is that their decoding can be reduced to the decoding of the smaller codes. This leads to a dramatic decrease in decoding complexity in overloaded cases. Additionally, we show that this method can be used for enlarging BWBE codes and arriving at other BWBE codes or ABWBE codes. Using these enlarged codes, we simulated some CDMA systems with binary input and binary signatures that employ low computational optimum or sub-optimum decoders. An important result of these simulations is that, TSC criterion is not an appropriate figure of merit when the input alphabet is binary (which is a case that is most encouraging in the practice). Hence despite the fact that WBE codes are optimum when the input data vector is real or complex [1], they are not appropriate for systems with binary inputs.

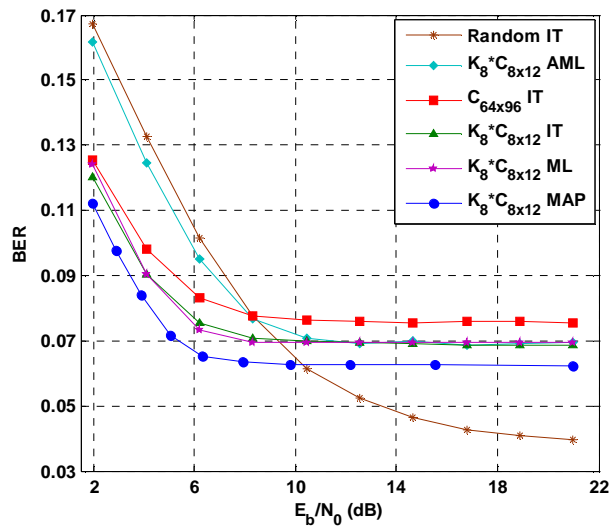

Fig. 3. BER versus $E_b/N_0$ for a system with 64 chips and 96 users using different codes and decoding methods. In the figure, * indicates the Kronecker product.

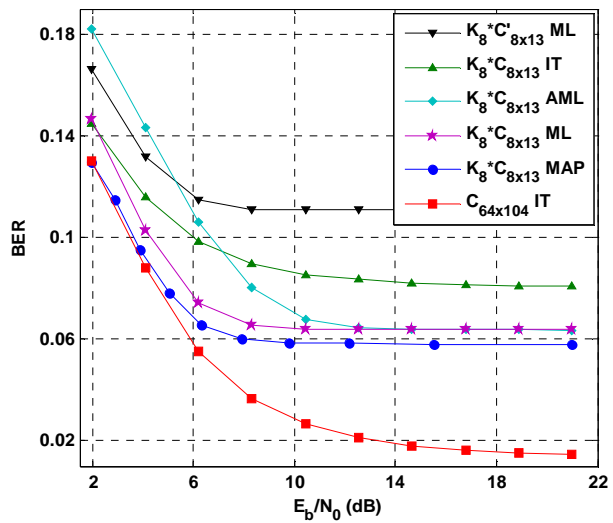

Fig. 4. BER versus $E_b/N_0$ for a system with 64 chips and 104 users using different codes and decoding methods. In the figure, * indicates the Kronecker product.